\documentstyle[prl,aps,floats,twocolumn,graphics,epsfig]{revtex}
\newcommand{\be}{\begin{equation}}
\newcommand{\bea}{\begin{eqnarray} \nonumber}
\newcommand{\ee}{\end{equation}}
\newcommand{\eea}{\end{eqnarray}}

 \def\(({\left(}
 \def\)){\right)}
\def\[[{\left[}
\def\]]{\right]}
\def\bi{\bibitem}
\def \form#1 {eq. (\ref{#1}) }
\def \parziale#1#2  {{\partial {#1} \over \partial {#2}}}

\def \ba#1 {\overline{#1}}

\def  \eps{\epsilon}

\title{Local fluctuation dissipation relation}
\author{ Giorgio Parisi }

\address{
  Dipartimento di Fisica, INFM, SMC and INFN,
Universit\`a di Roma {\em La Sapienza}, P. A. Moro 2, 00185 Rome, Italy. }

\date{\today}

\begin{document}
\twocolumn[ \hsize\textwidth\columnwidth\hsize\csname
@twocolumnfalse\endcsname
\maketitle

\begin{abstract}

In this letter I show that the recently proposed local version of the fluctuation dissipation relations follows from the
general principle of stochastic stability in a way that is very similar to the usual proof of the fluctuation
dissipation theorem for intensive quantities.  Similar arguments can be used to prove that all sites in an aging
experiment stay at the same effective temperature at the same time.
 \end{abstract}

\pacs{75.10.Nr, 75.40.Mg, 02.60.Pn}
]

The fluctuation dissipation relations in off-equilibrium dynamics are a crucial tool to explore the landscape of a
disordered system.  These fluctuation dissipation relations are different from the predictions of the fluctuation
dissipation theorem at equilibrium, so that sometimes one speaks about violations of the fluctuation dissipation
theorem.  The existence of these new relations has been firstly proved in mean field theories
\cite{CUKU,framez,CKPA,revaging}, however it has been later shown that they follow from the general principle of
stochastic stability \cite{GUERRA,AI,SOL,NOI}, that roughly speaking asserts that a random perturbation acts in a smooth
way.

The fluctuation dissipation relations can be expressed in a rather simple form that can be easily interpreted from the
theoretical point of view using the concept of the effective temperature \cite{CKP,FV}.  Moreover the main parameters
entering into the fluctuation dissipation relations have a simple interpretation from the point of view of equilibrium
statistical mechanics \cite{CUKU,framez,NOI}.  This fact implies that for a given system the form of the
fluctuation dissipation relations is universal in off-equilibrium dynamics, i.e. it does not depend on the way in which
the system is put in an off-equilibrium situation (as soon the system remains slightly out of equilibrium).
These fluctuation dissipation relations have been amply observed in numerical simulations
\cite{FR,RM,P,K} and quite recently in real experiments \cite{IG,BC,HO}.

The case that has been mostly studied is where we consider observables that are the average over the whole sample.  In
this case there is a static-dynamic relation that connects the dynamic fluctuation dissipation relations to the static
average of global quantities.  Recent there have been a few investigations on fluctuation dissipation relations that
involve only given local variables \cite{CC,MR}.  The aim of this letter is to give a theoretical foundation to these
local fluctuation dissipation relations and to derive the appropriate static-dynamic relation.  In order to reach this
goal we have to introduce a new concept: the probability distribution of the single spin overlap: this is a
generalization of the well know probability distribution of the total overlap of the system.

Let us recall the usual equilibrium fluctuation theorem.  If we consider a pair of conjugated variables (e.g. the
magnetic field and the magnetization) the response function and the spontaneous fluctuations are
deeply related.  Indeed, if $R_{eq}(t)$ is the integrated response (i.e. the variation of the magnetization at time $t$
when we add a a magnetic field from time 0 on) and $C_{eq}(t)$ is the correlation among the magnetization at time zero
and at time $t$, we have that $ R_{eq}(t)=\beta (C_{eq}(0) -C_{eq}(t)) $, where $\beta = (kT)^{-1}$.  If we we eliminate
the time and we plot parametrically $R_{eq}$ as function of $C_{eq}$, we have that
\begin{equation}
-{ d R_{eq} \over dC_{eq} }=\beta \ .
\end{equation}
The previous relation can be considered as the definition of the temperature and it is a consequence of the zeroth law
of the thermodynamics.

In an aging system the generalized fluctuation dissipation relations can be formulated as follows.  Let us suppose that
the system is carried from high temperature to low temperature at time 0 and it is in an aging regime.  We can define a
response function $R(t_{w},t)$ as the variation of the magnetization at time $t$ when we add a a magnetic field from
time $t_{w}$ on.  In a similar way $C(t_{w},t)$ is the correlation among the magnetization at time $t_{w}$ and at time
$t$.  We can define a function $R_{t_{w}}(C)$: we must plot $R(t_{w},t)$ versus $C(t_{w},t)$ by eliminating the time $t$
in the region $t>t_{w}$, where the response function is different from zero.

The fluctuation dissipation relations state that for large $t_{w}$ the function $R_{t_{w}}(C)$ converge to a limiting
function $R(C)$.  We can define
\begin{equation}
-{ d R \over dC} =\beta X(C)
\end{equation}
where $X(C)=1$, for $C>C_{\infty}\equiv\lim_{\to \infty}C_{eq}(t)$, and $X(C)<1$ for 
$C<C_{\infty}$.  

The shape of the function $X(C)$ gives important information on the free energy landscape 
of the problem, as discussed at lengthy in the literature \cite{CUKU,framez,CKPA,revaging,GUERRA,AI,SOL,NOI}.  
It has been shown that in stochastically stable systems the function $X(C)$ is related to basic equilibrium properties of
the system.  Let us illustrate this point by considering for definitiveness the case of a spin glass.  Given two
equilibrium configurations $\sigma$ and $\tau$, we define the global overlap as
\begin{equation}
q(\sigma,\tau) ={\sum_{i=1,N}\sigma_{i}\tau_{i} \over N}\ ,
\end{equation}
where $N$ is the total number of spins.  The function $P_{J}(q)$ is the probability distribution of the overlap for a
given sample and the function $P(q)$ is defined as the average of $P_{J}(q)$ over the samples.

It is convenient to introduce the function $x(q)$ defined by $d P(q) / d q = x(q)$.
The  relation among the dynamic  fluctuation dissipation relations  and the statics quantities is simple
$
X(C)=x(C)
$.

There are recent numerical results \cite{CC,MR} that indicate that the fluctuation dissipation relations  and
the static-dynamics connection can be generalized to local quantities in systems where a quenched disorder is present
and aging is heterogeneous \cite{HE}.

For {\sl one given sample} we can consider the local integrated response function $R_{i}(t_{w},t)$, that is the
variation of the magnetization at the point $i$ at time $t$ when we add a magnetic field at the point $i$ starting at
the time $t_{w}$.  In a similar way the local correlation function $C_{i}(t_{w},t)$ is defined to be the correlation
among the spin at the point $i$ at different times (i.e. $t_{w}$ and $t$).  Quite often in system with quenched
disorder aging is very heterogenous: the function $C_{i}$ and $R_{i}$ change dramatically from one point to an other
\cite{HE}.

It has been observed in simulations \cite{CC,MR} that local fluctuation dissipation relations  seems to hold
\begin{equation}
-{ d R_{i} \over dC_{i} } =\beta X_{i}(C)\ ,
\end{equation}
where $X_{i}(C)$ has quite strong variations with the site $i$.

It has also been suggested that in spite of these strong heterogeneities, if we define the effective $\beta^{eff}_{i}$
at time $t$ at the site $i$ as
\begin{equation}
-{ d R_{i}(t_{w},t) \over dC_{i}(t_{w},t) } =\beta X_{i}(t_{w},t)\equiv\beta^{eff}_{i} 
(t_{w},t) \ ,
\end{equation}
the quantity $\beta^{eff}_{i}(t_{w},t)$ does not depend on the site \cite{MR}.  In other words a thermometer coupled to
a given site would measure (at a given time) the same effective temperature independently on the site: different sites
are thermometrically indistinguishable.

These empirical results call for a theoretical explanation.  The aim of this note is to show that these results are
consequence of stochastic stability in an appropriate contest and that there is a local relation among static and
dynamics.  The crucial step consists in defining a local probability distribution of the overlap for a given system at
point $i$ (i.e. $P_{i}(q)$). We will then introduce the notation $ x_{i}(q)=\int_{0}^{q}P_{i}(q')dq' \ $ and we
will show that the static-dynamic connection for local variables is very similar to the one for global variables and it
is given by $
X_{i}(C)=x_{i}(C) $.

It is evident that we need a definition of the local overlap that should be rather different from the usual one.  Indeed the local
overlap of two equilibrium configurations (i.e. $\sigma_{i}\tau_{i}$) is always equal to $\pm 1$. If we use a naive
definition, we find that the probability distribution of the local overlap is the sum of two delta functions at $\pm 1$.

An useful definition of the local overlap an of its probability distribution can be obtained if we consider $M$
identical copies (or clones) of our sample: we introduce $N \times M$ $\sigma^{a}_{i} $ variables where $a=1,M$
(eventually we send $M$ to infinity) and $N$ is the (large) size of our sample ($i=1,N$).  The Hamiltonian in this Gibbs
ensemble is just given by
\begin{equation}
H_{K}(\sigma)=
\sum_{a=1,M} H(\sigma^{a})  +\eps H_{R}[\sigma] \ , 
\end{equation}
where $H(\sigma^{a})$ is the Hamiltonian for a fixed choice of the couplings and the  $H_{R}[\sigma]$
is a random Hamiltonian that couples the different copies of the system.  A possible choice is 
\begin{equation}
    H_{R}[\sigma]=\sum_{a=1,M;i=1,N}K^{a}_{i}\sigma_{i}^{a}\sigma_{i}^{a+1} \ ,
\end{equation}
where
the variables $K^{a}_{i}$ are identically
distributed independent random Gaussian variables with zero average and variance 1.  We can consider other ways to
couple the systems (e.g. $ H_{R}[\sigma]=\sum_{a,b=1,M;i=1,N}K^{a,b}_{i}\sigma_{i}^{a}\sigma_{i}^{b}$)
but let us stick to the previous one for definitiveness.

Our central hypothesis is that all intensive self average quantities are smooth function of $\eps$ for small $\eps$. 
This hypothesis is a kind of generalization of stochastic stability.  According to this hypothesis, when $\eps \to
0$, the local correlation function and the response functions tend to the result for $\eps=0$  uniformly in time.

In the case of small, but non-zero $\eps$, let us take two equilibrium configurations $\sigma$ and $\tau$. For given 
$K$, we
consider  the site dependent overlap
\begin{equation}
q_{i}(\sigma.\tau)={\sum_{a=1,M} \sigma^{a}_{i} \tau^{a}_{i} \over M} \ .
\end{equation}
We define the $K$-dependent probability distribution $P^{K}_{i}(q)$ as the probability distribution of the previous
overlap.  By averaging over $K$ at fixed $\eps,$ we can define
\begin{equation}
P^{\eps}_{i}(q)=\overline{P^{K}_{i}(q)}\ ,
\end{equation}
where the bar denotes the average over $K$.
We finally define
\begin{equation}
P_{i}(q)=\lim_{\eps \to 0}P^{\eps}_{i}(q) \ ,
\end{equation}
where the limit $\eps \to 0$ is done {\sl after} the limits $M \to \infty$ and $N \to \infty$ (alternatively we keep
$\eps^{2} M$ and $\eps^{2} N$ much larger than 1).

The proof of the local fluctuation dissipation relations can be obtained by copying, {\sl mutatis mutandis}, the proof of
the usual fluctuation dissipation relations \cite{NOI}.  For example we can consider the following perturbation
 \begin{equation}
\Delta H^{(2)}_{i} \equiv \sum_{a=1,M,b=1,M} h^{a,b}\sigma^{a}_{i} \sigma^{b}_{i} \ ,
\end{equation}
where the variables $h$ are Gaussian random variables with zero average and variance $\delta/M$.  If we assume for
simplicity a Langevin type of time evolution, the same steps of \cite{NOI} give :
\begin{equation}
\chi^{(2)}_{i}(t)=2\int_{t_{w}}^{t}d t'C_{i}(t',t) {\partial R_{i}(t',t) \over \partial t'} \ .
\end{equation}
The previous equation can be rewritten as
\begin{equation}
\chi^{(2)}_{i}(t) =2 \int dC C\ X_{i}(t)(C) \ ,
\end{equation}
where we have defined
\begin{equation}
X_{i}(t)(C(t',t))={\partial R_{i}(t',t) \over \partial t'}\left({\partial C_{i}(t',t) \over \partial t'}\right)^{-1} \ .
\end{equation}
If we make the crucial hypothesis  that the limits $t\to \infty $ and $\delta \to 0$ may be exchanged, we get
\begin{equation}
\lim_{t \to \infty}\chi^{(2)}_{i}(t)=\chi^{(2)}_{i}=\beta \int dq P_{i}(q) (1-q^{2}) \ .
\end{equation}

Generalizing the previous arguments we get
\bea
\chi^{(s)}_{i}(t)=s\int_{t_{w}}^{t}d t'C_{i}(t',t)^{s-1}{\partial R_{i}(t',t) \over \partial t'}=\\
s \int dC C^{s-1}X_{i}(t)(C)\to_{t \to \infty}
\beta \int dq P_{i}(q) (1-q^{s})=\\
s \beta\int d q \ x_{i}(q)q^{s-1} \nonumber
\eea
Using the previous equations for all values of $s$, we  arrive to the conclusion that  the quantity  $X_{i}(t)(C)$ has a 
limit when the time goes to infinity (i.e. to formulation of the local fluctuation dissipation relations). The 
limit is given by
\begin{equation}
X_{i}(C)=\lim_{t \to \infty} X_{i}(t)(C)=x_{i}(C)\equiv\int_{0}^{C} dqP_{i}(q) dq\ .
\end{equation}
The previous equations give the connection among  static quantities and the local fluctuation dissipation relations.

 The thermometric indistinguishability of the sites can be proved by considering two far away sites $i$ and $k$ and by
 introducing a perturbation that depends both the spins at $i$ and the spins at $k$.

A typical example is
\begin{equation}
\Delta H^{(3,2)}_{i,k} \equiv \sum_{a_{1},a_{2},a_{3},b_{1},b_{2}=1,M} h^{a_{1},a_{2},a_{3},b_{1},b_{2}}_{i}
\sigma^{a_{1}}_{i} \sigma^{a_{2}}_{i} \sigma^{a_{3}}_{i} \sigma^{b_{1}}_{i} \sigma^{b_{2}}_{i} 
\end{equation}
where the variables $h$ are Gaussian random variables with zero average and variance $\delta/M^{4}$.  If we proceed as
before, we find that the probability distribution $P_{i,k}(q_{i},q_{k})$ can be written in terms of $X_{i}(t)$ and
$X_{k}(t)$.  By imposing that $P_{i,k}(q_{i},q_{k})$ is  a positive distribution, we finds that
\begin{equation}
X_{i}(t)=X_{k}(t) \ . \label{PRIMA}
\end{equation}
This conditions is just what is needed to impose the thermometric indistinguishability of the sites during aging.

As a by product we obtain
\begin{equation}
P_{i,k}(q_{i},q_{k})=\int_{0}^{1}dx\delta(q_{i}-q_{i}(x))\delta(q_{k}-q_{k}(x)) \label{SECONDA}
\end{equation}
where $q_{i}(x)$ is the inverse function of $x_{i}(q)$.  This relation has some interesting consequences that will be
investigated in details elsewhere.

We note that the probability distribution of the local overlap $P_{i}(q)$, being related to a dynamical quantity,  must 
depend only on the local environment around the point $i$ and therefore it must have a straightforward limit when the 
volume of the system goes to infinity (e.g it should be independent of the boundary conditions). It is remarkable that for 
far away points ($i,k$) the two probability distributions   $P_{i}(q)$ and $P_{k}(q)$ are independent one from the 
other, but the joint probability distribution of $q_{i}$ and $q_{k}$ does not factorize has shown by eq. \ref{SECONDA}. 

Summarizing, we have seen that for a given sample it is possible to give a definition of a local probability
distribution of the overlap, that depend on the site.  The properties of this local probability distribution are related
to the local fluctuation dissipation relations, that automatical follows from the present formalism.  The condition of
thermometric indistinguishability of the sites turns out to be a byproduct of our approach: during the aging regime all
the sites are characterized by the same effective temperature.

It is important to stress that here (as well as in the original paper \cite{NOI}) in the all the derivations we have made
no esplicite assumption concerning the validity of static or dynamical ultrametricity or on the existence of a large
separations of the time scales.  Although the generalized fluctuation dissipation relations have been firstly derived in
ultrametric mean field models \cite{CUKU,framez}, their validity relies only on more general properties like stochastic stability.

I am grateful to Jorge Kurchan for an illuminating discussion.

\end{document}